\documentclass[showpacs,amsmath,amssymb,prd,epfs,twocolumn,multicol]{revtex4}
\tighten
\usepackage{graphicx}
\usepackage{dcolumn}
\usepackage{bm}
\usepackage{epsf}
\usepackage{hyperref}
\usepackage{amsfonts}

\begin{document}

\title{Comment on ``High--order 
contamination in the tail of gravitational collapse''}

\author{Alexander Z.~Smith$^1$ and Lior M.~Burko$^{1,2}
\footnote{Email: burko@uah.edu}$}

\affiliation{$^1$ Department of Physics and
Astronomy, Bates College, Lewiston, Maine 04240 \\
$^2$ Department of Physics, University of Alabama in Huntsville,
Huntsville, Alabama 38599} 

\date{October 3, 2005}

\begin{abstract}
We confront the predictions of S.~Hod, Phys.~Rev.~D {\bf 60}, 104053 (1999) for the late--time decay rate of black 
hole perturbations with numerical data. Specifically, we ask two questions: First, are corrections to the Price tail 
dominated by logarithmic terms, as predicted by Hod? Second, if there were logarithmic correction terms, do they take the specific 
form predicted in Hod's paper? The answer to both questions is ``no.'' 
\end{abstract}

\pacs{04.70.Bw, 04.25.Nx, 04.30.Nk }

\maketitle

The late--time tail of black hole perturbations has been studied in detail  
\cite{price,leaver,ching,gundlach-et-al,burko-ori,burko-khanna,scheel,price-burko}. 
The leading--order (in inverse--time) behavior of the tail is well understood: it is given by a 
power--law in inverse--time, the power index depending on the multipole order of the initial perturbation. 
Specifically, the late--time 
behavior is given asymptotically by $t^{-m}$. For initial data with compact support, $m=2\ell+3$, $\ell$ being 
the multipole order. (For a rotating black hole mode coupling can generate lower values of $m$ if not 
disallowed \cite{burko-khanna}.)  
In Ref.~\cite{hod}, the question of higher--order corrections is addressed. Specifically, 
Ref.~\cite{hod} argues that the correction is dominated by a logarithmic term. Indeed, 
{\em naive} expectations support 
this prediction: as the effective potential that scatters the waves includes a logarithmic term (when expressed in 
``natural'' coordinates, such as the Regge--Wheeler ``tortoise'' coordinate), it is natural to expect that 
logarithmic terms would appear also in the scattered waves. 
Notwithstanding these expectations, in this Comment we show that the  
correction terms show no evidence of a logarithmic dependence, and that it is dominated by 
a simple power series. In fact, 
Ching {\em et al} \cite{ching} showed that the Schwarzschild potential is exceptional, in that the 
leading behavior is not logarithmic, 
although the potential is. Most logarithmic potentials would induce logarithmic tails. In this 
Comment we argue, and demonstrate 
numerically, that the conclusions of Ching {\em et al} apply not just to the leading order tail, 
but also to the leading 
subdominant term.

That simple higher-order power-series terms exist in the tail is a trivial observation: It is well known that the leading 
order term in an inverse--time expansion is given by $t^{-m}$. A small shift in the origin of time, i.e., 
the transformation $t\to t-t_0$  ($t\gg t_0$, as we are interested in the late--time limit, as 
$t\to\infty$) will result in an asymptotic behavior given by 
$(t-t_0)^{-m}\sim \,t^{-m} + m\,t_0\,t^{-(m+1)}$. Namely, a simple shift in the origin of time gives 
rise to a simple power series expansion of the late--time tail, that has {\em nothing} to do with black hole 
scattering. In actual simulations, the origin of time is typically chosen arbitrarily. That is, as no special 
attention is typically given to the question of where to place the origin of time, one would expect generic 
simulations to always include such simple--series terms. The interesting question then is not whether simple--series 
terms exist, but rather whether they are dominated by logarithmic terms, as predicted by Ref.~\cite{hod}. We expect that 
the answer to the latter question is ``no.'' These expectations are supported by a failure to reproduce logarithmic terms for the 
late--time tail, both by the present authors and by others  \cite{poisson-comment}.

In the picture that we present in this Comment, the effect of black--hole scattering would then be to add to the correction term, i.e., 
the behavior of the late--time field would now be given by 
$ t^{-m} + (a + m\,t_0)\,t^{-(m+1)}$, where $a$ is a constant (that may depend on 
the location of the evaluation point and the parameters of the pulse). Distinguishing between a pure power series and a power series that 
includes logarithmic terms is not easy. Specifically, it is not easy to find from the behavior of the field itself whether logarithmic 
correction terms appear or not, for the simple reason that the function $\ln t$ increases only very slowly with $t$, and 
there are practical limits to how long one can evolve the data numerically. Instead, it is useful to calculate auxiliary quantities---some of 
which have been introduced before---and look for such quantities for which the signature of the 
logarithmic correction terms is easily recognizable. In this Comment we focus on three such quantities: 
(a) the local power index $n:=-t\,\dot{\psi}/\psi$ \cite{burko-ori}, (b) $\delta:=(n-n_{\infty})\,(t/M)/\ln (t/M)$ 
\cite{hod}, and (c) $\Delta:=(n-n_{\infty})\,(t/M)$, where $n_{\infty}:=n\,(t\to\infty)$.  Here we assume, without loss of generality, that 
the perturbation is that of a Schwarzschild black hole of mass $M$, and that the perturbing field is a spherical ($\ell =0$) scalar 
field, so as to make $n_{\infty}=3$. In what follows, we assign a subscript ``$N$'' to numerically evaluated data, and a subscript 
``$H$'' to quantities predicted by Ref.~\cite{hod}.

\begin{figure}[t]
\input epsf
\centerline{ \epsfxsize 9.0cm \epsfbox{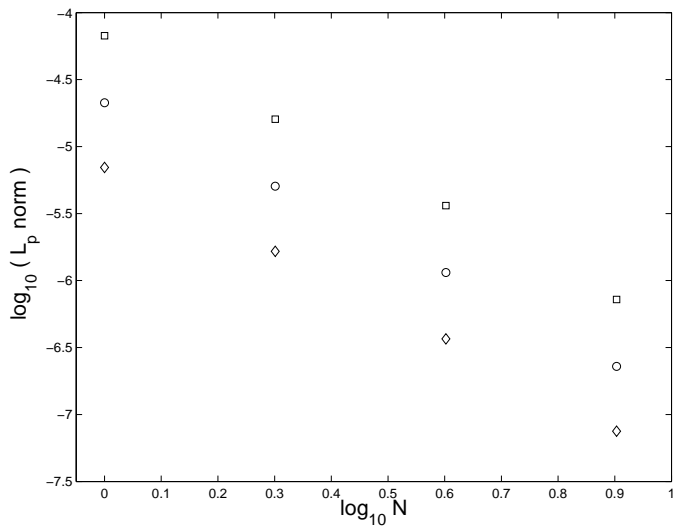}}
\caption{The H\"older $L_p$ vector norms for the local power index. Shown are the $L_p$ norms for $p=1,2,\infty$ for
a vector of ten equally spaced data points at $r_*=0$ between $t=9000M$ and $t=10000M$. Each norm corresponds to the vector
of differences of the data obtained for a grid density of $2N$ data points per $M$ and those obtained for a
grid density of $N$ data points per $M$. Data are shown for $N=1,2,4,8$. The $L_1$ (sum) norm is shown in squares,
the $L_2$ (Euclidean) norm in circles, and the $L_{\infty}$ (maximum) norm in diamonds.}
\label{fig0}
\end{figure}

\begin{figure}[t]
\input epsf
\centerline{ \epsfxsize 9.0cm \epsfbox{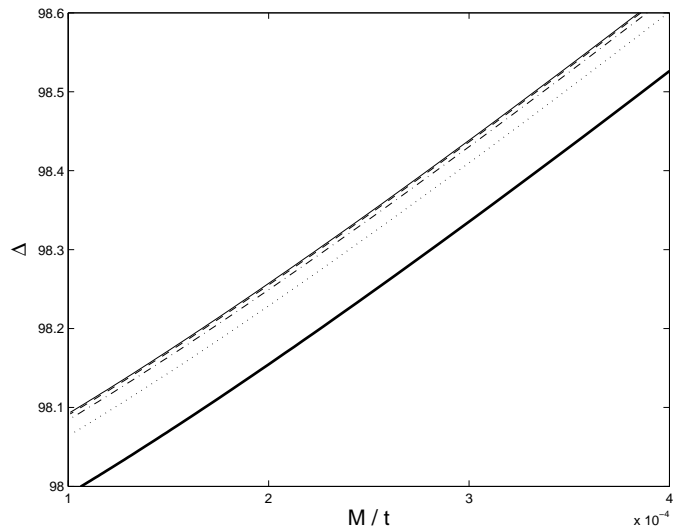}}
\caption{The diagnostic $\Delta$ as a function of $M/t$ for different grid densities 
(in terms of the number $N$ of grid point per $M$): Bold solid curve ($N=1$), 
dotted curve ($N=2$), dash--dotted curve ($N=4$), dashed curve ($N=8$), and thin solid curve ($N=16$).}
\label{fig0.5}
\end{figure}

\begin{figure}[t]
\input epsf
\centerline{ \epsfxsize 9.0cm \epsfbox{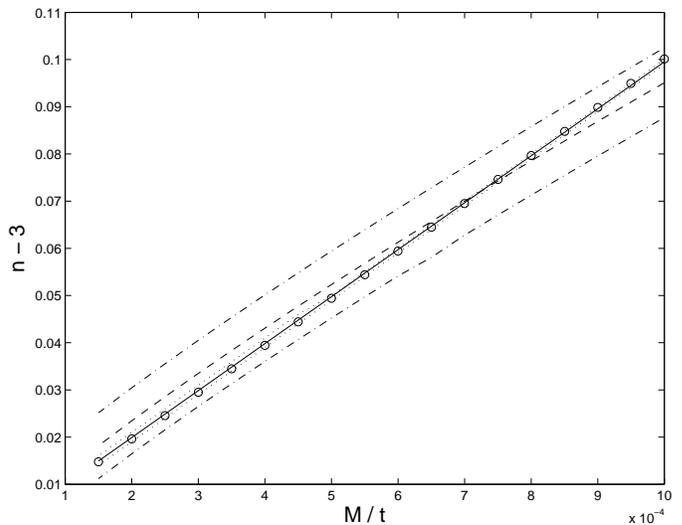}}
\caption{The local power index $n-3$ as a function of $M/t$. The numerical data points are depicted by circles; The solid
line is a best fit curve based on the ansatz (\ref{n-us}), and the two dotted curves are the $3\,\sigma$ confidence curves;  
The dashed curve is a best fit curve based on the ansatz (\ref{n-hod}), and the two dash--dotted curves are the
$3\,\sigma$ confidence curves.}
\label{fig1}
\end{figure}

We propose that the late--time behavior is given by a local power index $n$ as follows: 
\begin{equation}\label{n-us}
n = 3 +A\,\frac{M}{t}+B\,\frac{M^2}{t^2}+\cdots
\end{equation}
and the prediction of Ref.~\cite{hod} is that 
\begin{equation}\label{n-hod}
n_{H} = 3 + A_H\,\frac{\ln (t/M)}{t/M}+ B_H\,\frac{M}{t}+\cdots
\end{equation}
for some dimensionless constants $A,B;A_H,B_H$. In fact, Ref.~\cite{hod} predicts that $A_H=12$ exactly. One way to 
distinguish between the two predictions is to plot $n_N$ as a function of $M/t$, and test whether it 
approaches a straight line asymptotically as $M/t\to 0$. This is done in Fig.~\ref{fig1}, which displays the 
numerical data points and the best fit curves based on our ansatz (\ref{n-us}) and the ansatz of 
Ref.~\cite{hod} (\ref{n-hod}). The best 
fit is done using linear regression and the least squares method. We did not consider here the numerical noise in individual data 
points. We used a $1+1$D code in double--null coordinates whose convergence is second order. We specified initially incoming 
scalar--field perturbations on the characteristic hypersurface, with profile
$$\psi=\left[\frac{(v-v_1)(v_3-v)}{(v_2-v_1)(v_3-v_2)}\right]^8$$
for $v_1<v<v_3$, and $\psi=0$ otherwise, and with $v_2=(v_1+v_3)/2$. Here, $v$ is advanced time. In practice, we chose 
$v_1=10M$ and $v_3=30M$. Although the quantitative results (e.g., the value of the parameters $A,B, A_H,B_H$) depend on the 
choice of initial data, the qualitative behavior---and our conclusions---are independent of the choice of the initial data. The stability 
and second-order convergence of the code is demonstrated in Fig.~\ref{fig0}, that shows the H\"{o}lder $L_p$ vector norms of the 
local power index at various grid densities, and in Fig.~\ref{fig0.5} that shows the diagnostic $\Delta$ 
as a function of time for different grid densities. Other indicators behave similarly. However, all our diagnostics depend crucially on the local 
power index, whose calculation requires the numerical 
differentiation of the field. In fact, we found that the common double precision floating point 
arithmetic results in round off noise that prevents us from finding the phenomenon of interest easily. We solve the problem by using 
quadrupole precision floating point arithmetic. Our code exhibits clear second-order convergence globally, throughout the 
entire domain of the computation.

\begin{figure}
\input epsf
\centerline{ \epsfxsize 9.0cm \epsfbox{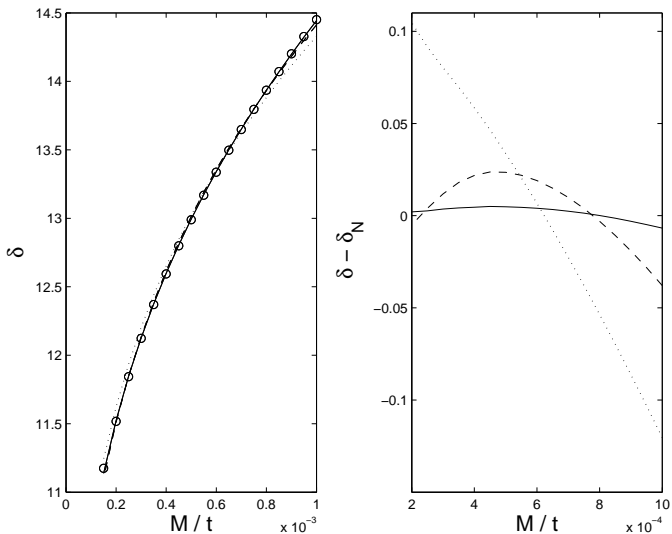}}
\caption{$\delta$ as a function of $M/t$. Left panel: The circles are the numerical data points; The dotted curve is a
first--order best fit curve, and the solid line a second--order best fit curve, based on  the ansatz (\ref{d1-us}); The
dashed curve is a second--order best fit curve based on the ansatz (\ref{d1-hod}). Right panel: The difference between the
three best fit curves and the numerical data.} \label{fig2}
\end{figure}

\begin{figure}[t]
\input epsf
\centerline{ \epsfxsize 9.0cm \epsfbox{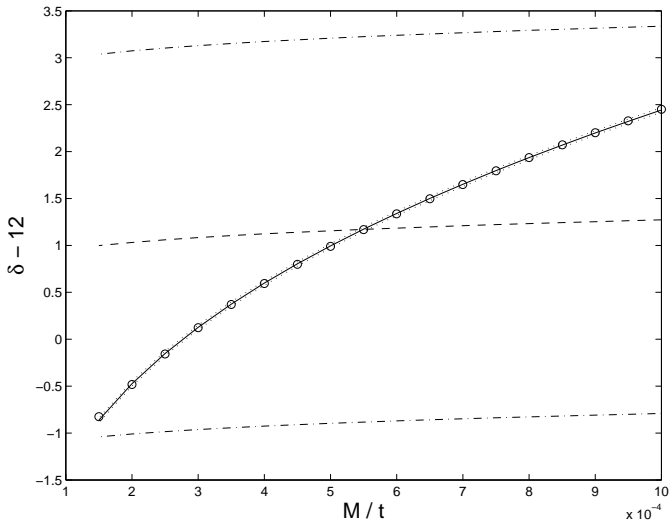}}
\caption{$\delta -12$ as a function of $M/t$. The circles are the numerical data points, the dashed line is a best fit curve
based on the ansatz (\ref{d1-hod}) with $A_H=12$, and the two dash--dotted curves are the $3\,\sigma$ confidence curves.
The solid curve is a best fit curved based on the ansatz (\ref{d1-us}) with the corresponding $3\,\sigma$ confidence
curves depicted by the dotted curves. }
\label{fig3}
\end{figure}

Figure \ref{fig1} displays the local power index as a funtion of time. It suggests that our ansatz fits the numerical data better than the ansatz of 
Ref.~\cite{hod}. Indeed, our curve has a corresponding squared correlation coefficient $R^2=0.99985$, whereas 
$R^2_H=0.98589$. 
Table \ref{table1} shows the best fit values for the parameters of the curves. Recall that not only does 
Ref.~\cite{hod} make the claim that there are logarithmic correction terms, it also predicts the value of the 
expansion parameter $A_H$ to equal $12$. However, the best fit finds that  $A_H=13.77\pm 0.17$, 
which deviates from the prediction of Ref.~\cite{hod} by over $10\,\sigma$. Although 
our best fit curves agrees with the numerical data to a greater extent, it is not easy to distinguish between 
an asymptotically linear curve (in $t^{-1}$), and a curve that behaves asymptotically like $\ln t/t$. In fact, 
one can construe the definition for $n$ as a differential equation for the field $\psi$:
\begin{equation}
\dot{\psi}+\frac{n(t)}{t}\,\psi=0\, ,
\label{deq}
\end{equation}
and if one can show that $n$ is asymptotically linear in $t^{-1}$, our ansatz is proved, and the prediction of 
Ref.~\cite{hod} falsified. However, as we have just seen, it is not easy to decide on this question from direct 
observation of $n$. We therefore consider other diagnostics. 

\begin{figure}[t]
\input epsf
?
\centerline{ \epsfxsize 9.0cm \epsfbox{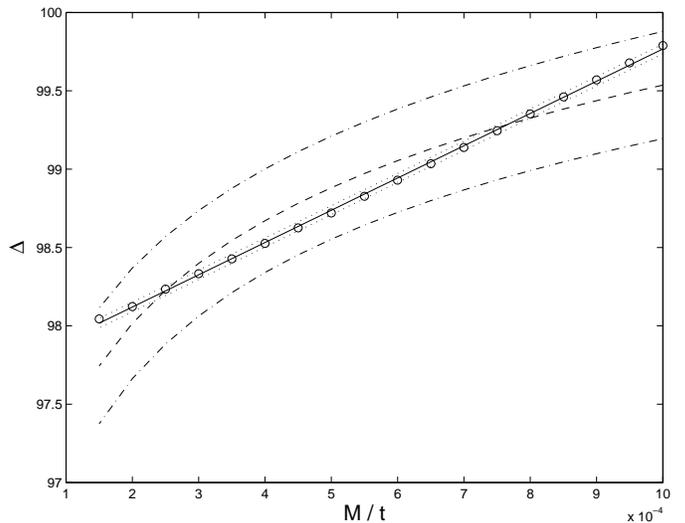}}
\caption{$\Delta$ as a function of $M/t$. The circles are the numerical data point, the solid curve is a best fit curve
based on the ansatz (\ref{d2-us}) with the corresponding $3\,\sigma$ confidence curves shown as the dotted curves, and the
dashed curve is the best fit curve based on the ansatz (\ref{d2-hod}), with the corresponding $3\,\sigma$ confidence curves 
shown by the dash-dotted curves.}
\label{fig4}
\end{figure}

\begin{table}[t]
\caption{Determination of the parameters $A,B$ and $A_H,B_H$, and the value
of the squared
correlation coefficients $R^2$ and $R^2_H$
for either ansatz, using the various diagnostics. The data above the single
horizontal line refers to our ansatz of no logarithmic terms, and the data
below the line refers to the prediction of Ref.~\cite{hod}. (a): In this
case, because the predicted $\Delta$ is constant, strictly speaking $R^2=0$.
(b): In this case $\Delta_H$ {\em anti}--correlates with the numerical
data; the former is a decreasing function of $M/t$, and the latter an
increasing function.
}
\begin{tabular}{llll}
Determination & & & \\
 from   &  $A$ or $A_H$  &      $B$ or $B_H$    &       $R^2$ or $R^2_H$     \\
\hline
\hline
$n-3$ (1st order)       &  $99.51\pm 0.13$ &  ---  &    $0.999851$ \\
$\delta$ (1st order)    &  $98.99\pm 0.12$ &  ---   &   $0.994955$ \\
$\delta$ (2nd order)    &  $97.76\pm 0.05$ & $2010\pm 73$ &   $0.999901$ \\
$\Delta$ (1st order)    &  $98.89\pm 0.13$ & --- & (a)  \\
$\Delta$ (2nd order)    &  $97.708\pm 0.008$ & $2059\pm 12$ & $0.99946$ \\
\hline
analytical & $12$ & --- & ---  \\
$n_H-3$ (1st order)     & $13.77\pm 0.17$ & ---  & $0.98589$ \\
$\delta_H$ (1st order) & $1.1\pm 0.23$ & --- & (a)  \\
$\delta_H$ (2nd order) & $-0.93\pm 0.09$ & $106.02\pm 0.67$ & $0.999365$ \\
$\delta_H-12$          & --- & $8.8\pm 1.7$ & $0.154808$ \\
$\Delta_H$ (1st order) & $12.96\pm 0.25$ & ---  & (b) \\
$\Delta_H$ (2nd order) & $-0.94\pm 0.07$ & $106.1\pm 0.5$ & $0.928968$
\label{table1}
\end{tabular}
\end{table}

The next diagnostic is $\delta$, which was first introduced in Ref.~\cite{hod}. Our prediction is that 
\begin{equation}\label{d1-us}
\delta = \frac{A}{\ln (t/M)}+\frac{B}{(t/M)\,\ln (t/M)}+\cdots
\end{equation}
whereas
\begin{equation}\label{d1-hod}
\delta_{H} = A_H+\frac{B_H}{\ln (t/M)}+\cdots\, .
\end{equation}
Figure \ref{fig2} shows $\delta$ (left panel) and $\delta - \delta_N$ (right panel) as functions of $M/t$, for 
our first-- and second--order predictions, and the second--order prediction of Ref.~\cite{hod}. As in the case 
of $n-3$, our ansatz appears to agree better with the numerical data. Indeed, we find from our first--order 
curve that $R^2=0.994955$ and from the second--order curve $R^2=0.999901$. The corresponding values for 
the predictions of Ref.~\cite{hod} are $R^2_H=0$ and 
$R^2_H=0.999365$. While the first--order prediction of Ref.~\cite{hod} fails spectacularly, its second-order 
prediction fits the numerical data quite well. (This, in itself, is not surprising, as the second--order term of 
Ref.~\cite{hod} is {\em identical} with our first--order term.) 
Also, the best fit determination for the parameters of the 
curve does not appear to converge for the predictions of Ref.~\cite{hod}: 
the first--order determination is $A_H=1.1\pm 0.2$ and 
the second--order determination is $A_H=-0.93\pm 0.09$. Recall that Ref.~\cite{hod} predicts not only logarithmic terms, 
but also predicts $A_H=12$. The latter result for $A_H$ deviates from the value of $12$ 
predicted in Ref.~\cite{hod} by $144\,\sigma$.   Next, we assume 
$A_H=12$, and present in Fig.~\ref{fig3} $\delta -12$ as a function of $M/t$. We find $R^2_H=0.154808$. Clearly, 
this prediction is not supported by the numerical data.

Lastly, we use $\Delta$ as a diagnostic. Our prediction for $\Delta$ is
\begin{equation}\label{d2-us}
\Delta = A+B\,\frac{M}{t}+\cdots
\end{equation}
whereas
\begin{equation}\label{d2-hod}
\Delta_{H} = A_H\,\ln (t/M)+B_H+\cdots\, .
\end{equation} 
Figure \ref{fig4} displays $\Delta$ as a function of $M/t$ for our second--order prediction, and the second--order
prediction of Ref.~\cite{hod}. Notice that the curvature of the curve predicted by Ref.~\cite{hod} has the wrong 
concavity (cf.~also with Fig.~\ref{fig0.5}). 
We find that $R^2=0.99946$, and $R^2_H=0.928968$. Also, $A_H=-0.94\pm 0.07$, which disagrees with the 
predicted value of $12$ by $198.2\,\sigma$. 

Our results hold also for the case of a Kerr black hole. 
Ref.~\cite{burko-khanna} has shown that indeed the local power 
index is asymptotically linear in $t^{-1}$ also for the case of Kerr perturbations. Using Eq.~(\ref{deq}), we conclude that the 
late--time field is a simple power series in inverse time also for the case of Kerr. (Notice that this question is separate 
from the question of mode couplings in the Kerr case.)  
Our results show that a simple power-series fits numerical data much better than an ansatz that include logarithmic terms. 
In addition, determination of the parameters using different diagnostics is inconsistent, when the prediction of 
Ref.~\cite{hod} is assumed. Moreover, the specific determination of Ref.~\cite{hod} that $A_H=12$ is shown here to be 
incorrect to a very high confidence level.

The authors thank Leor Barack for discussions. This work was funded in part by a grant 
to Bates College from the Howard Hughes Medical Institute.

\end{document}